\documentclass[preprint2]{aastex62}

\usepackage[normalem]{ulem}
\useunder{\uline}{\ul}{}
\usepackage{csquotes}
\usepackage{multirow}
\usepackage{amsmath}

\shorttitle{GRB 180728A}
\shortauthors{}

\begin{document}

\title{Two Predictions of supernova: GRB 130427A/SN 2013cq and GRB 180728A/SN 2018fip}

\correspondingauthor{R.~Ruffini}
\email{ruffini@icra.it}

\author{Y.~Wang}
\affiliation{ICRA and Dipartimento di Fisica, Sapienza Universit\`a di Roma, P.le Aldo Moro 5, 00185 Rome, Italy}
\affiliation{ICRANet, P.zza della Repubblica 10, 65122 Pescara, Italy}

\author{J.~A.~Rueda}
\affiliation{ICRA and Dipartimento di Fisica, Sapienza Universit\`a di Roma, P.le Aldo Moro 5, 00185 Rome, Italy}
\affiliation{ICRANet, P.zza della Repubblica 10, 65122 Pescara, Italy}
\affiliation{ICRANet-Rio, Centro Brasileiro de Pesquisas F\'isicas, Rua Dr. Xavier Sigaud 150, 22290--180 Rio de Janeiro, Brazil}

\author{R.~Ruffini}
\affiliation{ICRA and Dipartimento di Fisica, Sapienza Universit\`a di Roma, P.le Aldo Moro 5, 00185 Rome, Italy}
\affiliation{ICRANet, P.zza della Repubblica 10, 65122 Pescara, Italy}
\affiliation{Universit\'e de Nice Sophia Antipolis, CEDEX 2, Grand Ch\^{a}teau Parc Valrose, Nice, France}
\affiliation{ICRANet-Rio, Centro Brasileiro de Pesquisas F\'isicas, Rua Dr. Xavier Sigaud 150, 22290--180 Rio de Janeiro, Brazil}

\author{C.~Bianco}
\affiliation{ICRA and Dipartimento di Fisica, Sapienza Universit\`a di Roma, P.le Aldo Moro 5, 00185 Rome, Italy}
\affiliation{ICRANet, P.zza della Repubblica 10, 65122 Pescara, Italy}

\author{L.~Becerra}
\affiliation{Escuela de F\'isica, Universidad Industrial de Santander, A.A.678, Bucaramanga, 680002, Colombia}

\author{L.~Li}
\affiliation{ICRANet, P.zza della Repubblica 10, 65122 Pescara, Italy}

\author{M.~Karlica}
\affiliation{ICRA and Dipartimento di Fisica, Sapienza Universit\`a di Roma, P.le Aldo Moro 5, 00185 Rome, Italy}
\affiliation{ICRANet, P.zza della Repubblica 10, 65122 Pescara, Italy}
\affiliation{Universit\'e de Nice Sophia Antipolis, CEDEX 2, Grand Ch\^{a}teau Parc Valrose, Nice, France}



\begin{abstract}
On 2018 July 28, GRB 180728A triggered \textit{Swift} satellites and, soon after the determination of the redshift, we identified this source as a type II binary-driven hypernova (BdHN II) in our model. Consequently, we predicted the appearance time of its associated supernova (SN), which was later confirmed as SN 2018fip. A BdHN II originates in a binary composed of a carbon-oxygen core (CO$_{\rm core}$) undergoing SN, and the SN ejecta hypercritically accrete onto a companion neutron star (NS). From the time of the SN shock breakout to the time when the hypercritical accretion starts, we infer the binary separation $\simeq 3 \times 10^{10}$ cm. The accretion explains the prompt emission of isotropic energy $\simeq 3 \times 10^{51}$ erg, lasting $\sim 10$ s, and the accompanying observed blackbody emission from a thermal convective instability bubble. The new neutron star ($\nu$NS) originating from the SN powers the late afterglow from which a $\nu$NS initial spin of $2.5$ ms is inferred. We compare GRB 180728A with GRB 130427A, a type I binary-driven hypernova (BdHN I) with isotropic energy $> 10^{54}$ erg. For GRB 130427A we have inferred an initially closer binary separation of $\simeq 10^{10}$ cm, implying a higher accretion rate leading to the collapse of the NS companion with consequent black hole formation, and a faster, $1$ ms spinning $\nu$NS. In both cases, the optical spectra of the SNe are similar, and not correlated to the energy of the gamma-ray burst. We present three-dimensional smoothed-particle-hydrodynamic simulations and visualisations of the BdHNe I and II.
\end{abstract}

\section{Introduction} 
\label{sec:1}

By the first minutes of data retrieved from Konus-\textit{Wind}, \textit{Swift}, \textit{Fermi}, \textit{AGILE} or other gamma-ray telescopes \citep{1995SSRv...71..265A,2005SSRv..120..143B,2009ApJ...697.1071A,2009A&A...502..995T}, and the determination of redshift by VLT/X-shooter, Gemini, NOT or other optical telescopes \citep{1992A&A...257..811V,2004PASP..116..425H,2011A&A...536A.105V}, it is possible to promptly and uniquely identify to which of the nine (9) subclasses of gamma-ray bursts (GRBs) a source belongs (See table \ref{tab:GRBsubclasses} and the references therein). Consequently, it is possible to predict its further evolution, including the possible appearance time of an associated supernova (SN) expected in some of the GRB subclasses. This is what we have done in the case of GRB 130427A \citep{2015ApJ...798...10R, 2018ApJ...869..101R}, and in the present case of GRB 180728A.

GRB 130427A is a BdHN I in our model, details in section \ref{sec:bdhn} and in \citet{2014ApJ...793L..36F,2015PhRvL.115w1102F,2015ApJ...812..100B,2016ApJ...833..107B,2018ApJ...852..120B}. The progenitor is a tight binary system, of orbital period $\sim 5$~min, composed of a carbon-oxygen core (CO$_{\rm core}$), undergoing a SN event, and a neutron star (NS) companion accreting the SN ejecta and finally collapsing to a black hole (BH). The involvement of a SN in BdHN I and the low redshift of $z=0.34$ \citep{2013GCN..14455...1L,2013GCN..14478...1X,2013GCN..14491...1F} enable us to predict that the optical signal of the SN will peak and be observed $\sim 2$ weeks after the GRB occurrence at the same position of the GRB \citep{2013GCN..14526...1R}. Indeed the SN was observed \citep{2013GCN..14646...1D,2013ApJ...776...98X}. Details of GRB 130427A are given in section \ref{sec:130427A}.

The current GRB 180728A is a BdHN II in our model; it has the same progenitor as BdHN I, a binary composed of a CO$_{\rm core}$ and a NS companion, but with longer orbital period ($\gtrsim 10$~min), which is here determined for the first time. The CO$_{\rm core}$ undergoes SN explosion, the SN ejecta hypercritically accrete onto the companion NS. In view of the longer separation, the accretion rate is lower, it is not sufficient for the companion NS to reach the critical mass of BH. Since a SN is also involved in BdHN II and this source is located at low redshift $z=0.117$ \citep{GCN23055}, its successful prediction and observation were also possible and it is summarised in section \ref{sec:180728A_observation}. From a time-resolved analysis of the data in section \ref{sec:picture_and_data}, we trace the physical evolution of the binary system. For the first time we observed a $2$~s signal evidencing the SN shockwave, namely the emergence of the SN shockwave from the outermost layers of the CO$_{\rm core}$ \citep[see e.g.][]{1996snih.book.....A}. The SN ejecta expand and, after $10$~s, reach the companion NS inducing onto it a high accretion rate of about $10^{-3} M_\odot$~s$^{-1}$. Such a process lasts about $10$~s producing the prompt phenomena and an accompanying thermal component. The entire physical picture is described in section \ref{sec:picture_of_physics}, giving special attention to the new neutron star ($\nu$NS) originating from the SN. We explicitly show that the fast spinning $\nu$NS powers the afterglow emission by converting its rotational energy to synchrotron emission \citep[see also][]{2018ApJ...869..101R}, which has been never well considered in previous GRB models. We compare the initial properties of the $\nu$NS in GRB 130427A and in GRB 180728A, and derive that a $1$~ms $\nu$NS is formed in GRB 130427A while a $2.5$~ms $\nu$NS is formed in GRB 180728A. In section \ref{sec:neutron_star_charactristic} we relate the very different energetic of the prompt emission to the orbital separation of the progenitors, which in turn determines the spin of the $\nu$NS, and the rest-frame luminosity afterglows. We simulate the accretion of the SN matter onto the NS companion in the tight binaries via three-dimensional (3D) smoothed-particle-hydrodynamic (SPH) simulations \citep{2018arXiv180304356B}
that provide as well a visualisation of the BdHNe. The conclusions are given in section \ref{sec:conclusion}.

\section{Binary-driven hypernova}
\label{sec:bdhn}

Since the Beppo-SAX discovery of the spatial and temporal coincidence of a GRB and a SN \citep{1999A&AS..138..465G}, largely supported by many additional following events \citep{2006ARA&A..44..507W,2017AdAst2017E...5C}, a theoretical paradigm has been advanced for long GRBs based on a binary system \citep{2012ApJ...758L...7R}. It differs from the traditional theoretical interpretation of GRB which implicitly assumes that all GRBs originate from a BH with an ultra-relativistic jet emission \citep[see, e.g.,][]{1999PhR...314..575P,2004RvMP...76.1143P,Meszaros2002,Meszaros2006,2014ARA&A..52...43B,2015PhR...561....1K}.

Specifically, the binary system is composed by a CO$_{\rm core}$ and a NS companion in tight orbit. Following the onset of the SN, a hypercritical accretion process of the SN ejecta onto the NS occurs which markedly depends on the binary period of the progenitor \citep{2014ApJ...793L..36F,2015PhRvL.115w1102F,2015ApJ...812..100B,2016ApJ...833..107B}. For short binary periods of the order of $5$~min the NS reaches the critical mass for gravitational collapse and forms a BH \citep[see e.g.][]{2014A&A...565L..10R,2017arXiv171205001R,2018ApJ...852...53R}. For longer binary periods, the hypercritical accretion onto the NS is not sufficient to bring it to the critical mass and a more massive NS (MNS) is formed. These sources have been called BdHNe since the feedback of the GRB transforms the SN into a hypernova (HN) \citep{2017arXiv171205001R}. The former scenario of short orbital period is classified as BdHN type I (BdHN I), which leads to a binary system composed by the BH, generated by the collapse of the NS companion, and the $\nu$NS generated by the SN event. The latter scenario of longer orbital period is classified as BdNH type II (BdHN II), which leads to a binary NS system composed of the MNS and the $\nu$NS. 

Having developed the theoretical treatment of such hypercritical process, and considering as well other binary systems with progenitors composed alternatively of CO$_{\rm core}$ and BH, to NS and white dwarf (WD), a general classification of GRBs has been developed; see \citet{2016ApJ...832..136R} and Table~\ref{tab:GRBsubclasses} for details. We report in the table estimates of the energetic, spectrum and different component of the prompt radiation, of the plateau, and all the intermediate phases, all the way to the final afterglow phase. The GRBs are divided in two main classes, the BdHNe, which cover the traditional long duration GRBs \citep{1993ApJ...405..273W,Paczynski:1998ey}, and the binary mergers, which are short-duration GRBs \citep{1986ApJ...308L..47G,1986ApJ...308L..43P,Eichler:1989jb}. There are currently nine subclasses in our model, the classification depends on the different compositions of the binary progenitors and outcomes, which are CO$_{\rm core}$ and compact objects as BH, NS, and WD.  The same progenitors are possible to produce different outcomes, due to the different masses and binary separations.


\begin{table*}
\scriptsize
\centering
\begin{tabular}{c|ccccccccc}
\hline
Class &   Type  & Previous  & Number & \emph{In-state}  & \emph{Out-state} & $E_{\rm p,i}$ &  $E_{\rm iso}$  &  $E_{\rm iso,Gev}$  \\
& & Alias & & & & (MeV) & (erg) &  (erg) 	\\	
\hline
Binary Driven & I  & BdHN  & $329$ &CO$_{\rm core}$-NS  & $\nu$NS-BH & $\sim0.2$--$2$ &  $\sim 10^{52}$--$10^{54}$ &    $\gtrsim 10^{52}$ \\
Hypernova & II & XRF & $(30)$ &CO$_{\rm core}$-NS    & $\nu$NS-NS & $\sim 0.01$--$0.2$  &  $\sim 10^{50}$--$10^{52}$ &    $-$ \\
(BdHN) & III  & HN & $ (19) $ &CO$_{\rm core}$-NS    & $\nu$NS-NS & $\sim 0.01$  &  $\sim 10^{48}$--$10^{50}$ &    $-$ \\
& IV   & BH-SN & $5$ & CO$_{\rm core}$-BH  & $\nu$NS-BH & $\gtrsim2$ &  $>10^{54}$ &   $\gtrsim 10^{53}$   \\
\hline
 & I & S-GRF & $18$ &NS-NS & MNS & $\sim0.2$--$2$ &  $\sim 10^{49}$--$10^{52}$  &  $-$ \\
Binary & II  & S-GRB  & $6$ &NS-NS & BH & $\sim2$--$8$ &  $\sim 10^{52}$--$10^{53}$ &   $\gtrsim 10^{52}$\\
Merger& III  & GRF  & $(1)$ &NS-WD & MNS & $\sim0.2$--$2$ &  $\sim 10^{49}$--$10^{52}$ & $-$\\
(BM)& IV  & FB-KN$^{\star}$ & $(1)$ &WD-WD & NS/MWD & $ < 0.2$ &  $< 10^{51}$  & $-$\\
& V   & U-GRB & $(0)$ &NS-BH & BH & $\gtrsim2$ &  $>10^{52}$ & $-$ \\
\hline
\end{tabular}
\caption{Summary of the GRB subclasses. This table is an updated version of the one presented in \citet{2016ApJ...832..136R,2018ApJ...859...30R}. We unify here all the GRB subclasses under two general names, BdHNe and BMs. Two new GRB subclasses are introduced; BdHN Type III and BM Type IV. In addition to the subclass name in ``Class'' column and ``Type'' column, as well as the previous names in ``Previous Alias'' column, we report the number of GRBs with known redshift identified in each subclass updated by the end of 2016 in ``number'' column (the value in a bracket indicates the lower limit). We recall as well the ``in-state'' representing the progenitors and the ``out-state'' representing the outcomes, as well as the the peak energy of the prompt emission, $ E_{\rm p,i}$, the isotropic gamma-ray energy, $E_{\rm iso}$ defined in the $1$~keV to $10$~MeV energy range, and the isotropic emission of ultra-high energy photons, $E_{\rm iso,Gev}$, defined in the $0.1$--$100$ GeV energy range. 
We can see from this last column that this GeV emission, for the long GRBs is only for the BdHN Type I and Type IV, and in the case of short bursts is only for BM Type II and, in all of them, the GeV emission has energy more than $10^{52}$~erg.\\
$^{\star}$ We here adopt a broad definition of kilonova as its name, a phenomenon which is 1000 times more luminous than a nova. A kilonova can be an infrared-optical counterpart of a NS-NS merger. In that case the transient is powered by the energy release from the decay of r-process heavy nuclei processed in the merger ejecta \citep[e.g.][]{1998ApJ...507L..59L,2010MNRAS.406.2650M,2013Natur.500..547T,2013ApJ...774L..23B}. FB-KN stands for fallback-powered kilonova. We have shown that a WD-WD merger produces an infrared-optical transient from the merger ejecta, a kilonova, peaking at $\sim 5$ days post-merger but powered in this case by accretion of fallback matter onto the merged remnant \citep{2018JCAP...10..006R,2018arXiv180707905R}.  }
\label{tab:GRBsubclasses}
\end{table*}

\section{GRB 130427A as BdHN I} 
\label{sec:130427A}

GRB 130427A, as a BdHN I in our model, has been studied in our previous articles \citep{2015ApJ...798...10R,2018ApJ...869..101R}. This long GRB is nearby ($z=0.314$) and energetic ($E_{iso} \sim 10^{54}$~erg) \citep{2013GCN..14455...1L,2013GCN..14478...1X,2013GCN..14491...1F,2014Sci...343...48M}. It has overall the most comprehensive data to date, including the well observed $\gamma$-ray prompt emission \citep{2013GCN.14473....1V,GCN14487}, the full coverage of X-ray, optical and radio afterglow \citep{2013ApJ...779L...1K,2014ApJ...781...37P,2014Sci...343...38V,2014MNRAS.444.3151V,2014ApJ...792..115L,2014MNRAS.440.2059A,2017ApJ...837..116B}, and the long observation of the ultra-high energy emission (UHE) \citep{2013ApJ...771L..13T,2014Sci...343...42A,2015ApJ...800...78A}. Also it has been theoretically well-studied, involving many interpretations, including:  a black hole or a magnetar as the central engine \citep{2014MNRAS.439L..80B}; an unaccountable temporal spectral behaviors of the first $2.5$~s pulse by the traditional models \citep{2014Sci...343...51P}; the reverse-forward shock synchrotron model and its challenges in explaining the afterglow \citep{2013ApJ...776..119L,2016ApJ...818..190F,2016MNRAS.462.1111D,2017Galax...5....6D}; the synchrotron or the inverse Compton origins for the ultra-high energy photons \citep{2013ApJ...773L..20L,2013ApJ...776...95F,2013MNRAS.436.3106P,2014IJMPS..2860174T,2014ApJ...789L..37V}; the missing of the neutrino detection and its interpretation \citep{2013ApJ...772L...4G,2016MNRAS.458L..79J}. Our interpretation is alternative to the above traditional approach: 1) long GRBS are traditionally described as single systems while we assume a very specific binary systems as their progenitors. 2) The roles of the SN and of the $\nu$NS are there neglected, while they are essential in our approach as evidenced also in this article. 3) A central role in the energetics is traditionally attributed to the kinetic energy of ultra-relativistic blast waves extending from the prompt phase all the way to the late phase of the afterglow, in contrast to model-independent constraints observed in the mildly relativistic plateau and afterglow phases \citep{2015ApJ...798...10R,2018ApJ...869..151R,2018ApJ...869..101R}. In our approach the physics of the $e^+e^-$ plasma and its interaction with the SN ejecta as well as the pulsar-like behaviour of the $\nu$NS are central to the description from the prompt radiation to the late afterglow phases \citep{2018ApJ...852...53R}. One of the crucial aspects in our approach is the structure of the SN ejecta which, under the action of the hypercritical accretion process onto the NS companion and the binary interaction, becomes highly asymmetric. Such a new morphology of the SN ejecta has been made possible to be visualized thanks to a set of three-dimensional numerical simulations of BdHNe \citep{2014ApJ...793L..36F,2016ApJ...833..107B,2018arXiv180304356B}.

On this ground, soon after the observational determination of the redshift \citep{2013GCN..14455...1L}, by examining the detailed observations in the early days, we identified the BdHN origin of this source. On 2013 May 2, we made the prediction of the occurrence of SN 2013cq on GCN \citep[][quoted in appendix \ref{sec:gcns}]{2013GCN..14526...1R}, which was duly observed in the optical band on 2013 May 13 \citep{2013GCN..14646...1D,2013ApJ...776...98X}. 

To summarize  our work on this GRB: in \citet{2015ApJ...798...10R} we presented the multiwavelength light curve evolution and interpreted them by a tight binary system with orbital separation $\sim 10^{10}$~cm. GRB 130427A has a very bright prompt $\gamma$-ray spike in the first $10$~s, then it decays, coinciding with the rising of the UHE (100 MeV$−-$100 GeV) emission. The UHE peaks at $\sim 20$~s, then gradually dims for some thousand seconds. Soft X-ray observations start from $195$~s, it has a steep decay then follows a normal power-law decay $\sim t^{-1.3}$. We evidenced the presence of a blackbody component in the soft X-ray data in the time-interval from $196$~s to $461$~s; within which the temperature decreases from $0.5$~keV to $0.1$~keV. The thermal component indicates an emitter expanding from $\sim 10^{12}$~cm to $\sim 10^{13}$~cm with velocity $\sim 0.8~c$. This mildly relativistic expansion from our model-independent inference contrasts with the traditional ultrarelativistic external shockwave interpretation \citep[see e.g.][]{1998ApJ...497L..17S}. We attributed this thermal emission to the transparency of the SN ejecta outermost layer after being heated and accelerated by the energetic $e^+e^-$ plasma outflow of the GRB. The numerical simulations of this hydrodynamics process were presented in \citet{2018ApJ...852...53R}. As it is shown there, the resulting distance, velocity, and occurring time of this emission are all in agreement with the observations. Later in \citet{2018ApJ...869..101R}, we showed that the mildly relativistic ejecta can also account for the nonthermal component, in the early thousands of seconds powered by its kinetic energy, and afterward powered by the release of rotational energy of the millisecond-period $\nu$NS via a pulsar-like mechanism. The synchrotron emission well reproduces the observed optical and X-ray afterglow. A similar application of the $\nu$NS on GRB 180728A will be presented in section \ref{sec:afterglow_from_pulsar}, as well as the comparison to GRB 130427A.

\section{Observation and Prediction} 
\label{sec:180728A_observation}

On 2018 July 28, we had the opportunity to make a prediction of the SN appearance in a BdHN II. 

At 17:29:00 UT, On 2018 July 28, GRB 180728A triggered the \textit{Swift}-BAT. The BAT light curve shows a small precursor and $\sim 10$~s later it was followed by a bright pulse of $\sim 20$~s duration \citep{GCN23046}. \textit{Swift}-XRT did not slew to the position immediately due to the Earth limb, it began observing $1730.8$~s after the BAT trigger \citep{GCN23049}. The \textit{Fermi}-GBM triggered and located GRB 180728A at 17:29:02.28 UT. The initial \textit{Fermi}-LAT bore-sight angle at the GBM trigger time is 35 degrees, within the threshold of detecting GeV photons, but no GeV photon was found. The GBM light curve is similar to the one of Swift-BAT, consisting of a precursor and a bright pulse, the duration ($T_{90}$) is about $6.4$~s ($50$--$300$~keV) \citep{GCN23053}. A red continuum was detected by VLT/X-shooter and the absorption features of Mg II (3124, 3132), Mg I (3187), and Ca II (4395, 4434) were consistent with a redshift of $z=0.117$ \citep{GCN23055}.


After the detection of the redshift, On 2018 July 31, we classify this GRB as an BdHN II in our model, based on its duration, peak energy, isotropic energy, and the existence of photons with energy $> 100$~MeV, criteria in table \ref{tab:GRBsubclasses}. BdHN II involves the Type Ib/c supernova phenomenon, therefore, we predicted that a SN would appear at $14.7\pm 2.9$~days \citep{GCN23066} and be observed due to its low redshift. On 18 Augest 2018, \citet{GCN23142} on behalf of the VLT/X-shooter team reported the discovery of the SN appearance, which was confirmed in \citet{GCN23181}. The text of these GCNs are reported in the appendix \ref{sec:gcns}. The SN associated with GRB 180728A was named as SN 2018fip. Our prediction was confirmed.

We also predicted the supernova appearance in GRB 140206A \citep{2014GCN.15794....1R} and GRB 180720A \citep{GCN23019}, but unfortunately the optical observation does not cover the expected time ($\sim 13$ days after the GRB trigger time) of the supernova appearance.

\section{Data Analysis} 
\label{sec:picture_and_data}

\begin{figure*}
\centering
\includegraphics[width=0.8\hsize,clip]{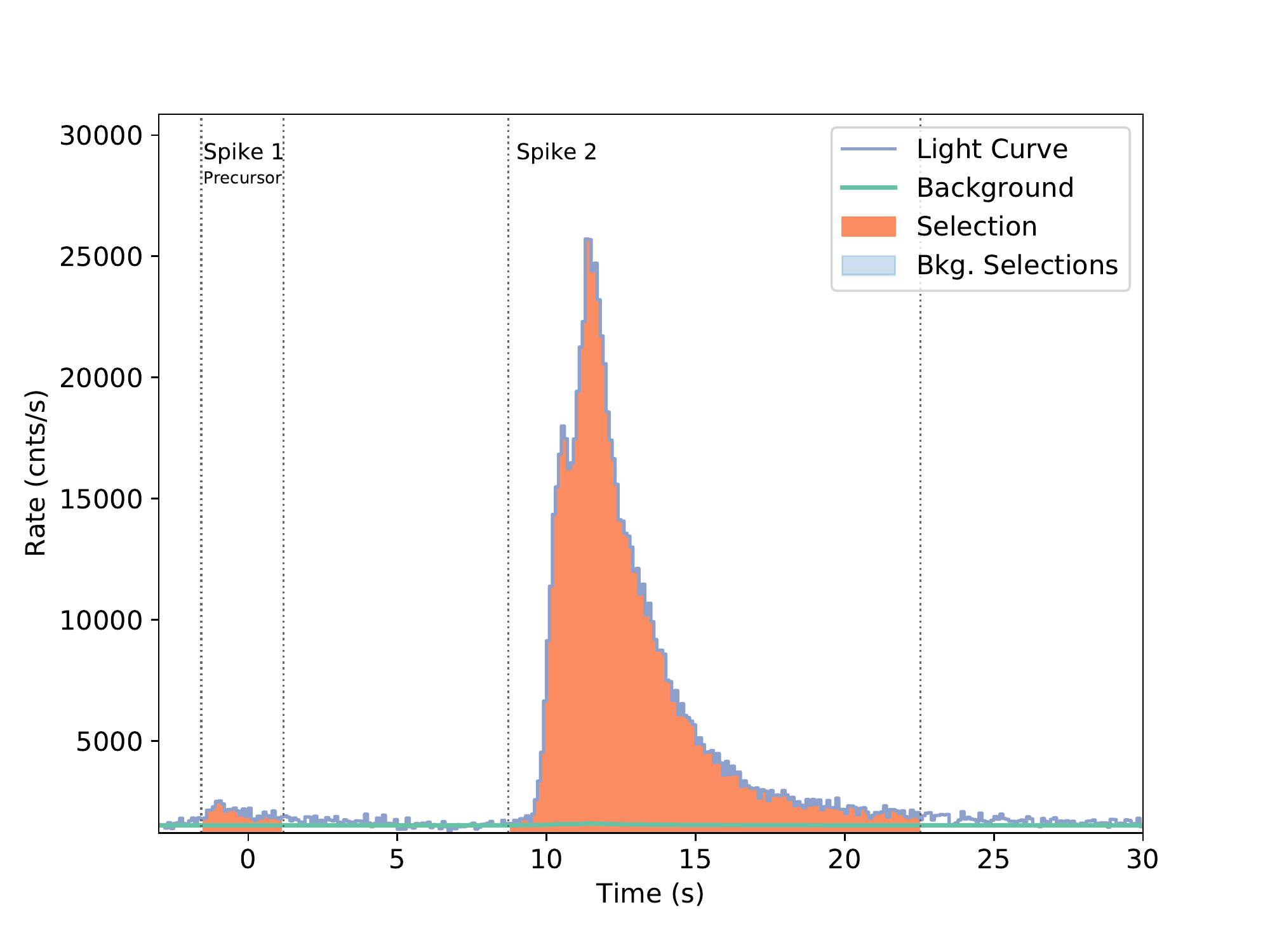}
\caption{Count rate light curve of the prompt emission: Data are retrieved from the NaI7 detector on-board Fermi-GBM. The prompt emission of GRB 180728A contains two spikes. The first spike, the precursor, ranges from $-1.57$~s to$1.18$~s. The second spike, which contains the majority of energy, rises at $8.72$~s, peaks at $11.50$~s, and fades at $22.54$~s.}
\label{fig:prompt_all_180728A}
\end{figure*}

GRB 180728A contains two spikes in the prompt emission observed by \textit{Swift}-BAT, \textit{Fermi}-GBM and Konus-\textit{Wind} \citep{GCN23046,GCN23053,GCN23055}. In the following we defined our $t_0$ based on the trigger time of \textit{Fermi}-GBM. The first spike, we name it as precursor, ranges from $-1.57$~s to $1.18$~s. And the second spike, which contains the majority of energy, rises at $8.72$~s, peaks at $11.50$~s, and fades at $22.54$~s, see figure~\ref{fig:prompt_all_180728A}. These time definitions are based on the count rate light curve observed by \textit{Fermi}-GBM, and determined by applying the Bayesian block method \citep{1998ApJ...504..405S}. \textit{Swift}-XRT started to observe $1730.8$~s after the BAT trigger, the luminosity of the X-ray afterglow follows a shallow decay with a power-law index $-0.56$ till $\sim 5000$~s, then a normal decay with a power-law index $-1.2$, which is a typical value \citep{2015ApJ...805...13L,2018ApJS..234...26L}.

\subsection{Prompt Emission: Two Spikes} 
\label{sec:first_spike}

\begin{figure*}
\centering
\includegraphics[width=0.48\hsize,clip]{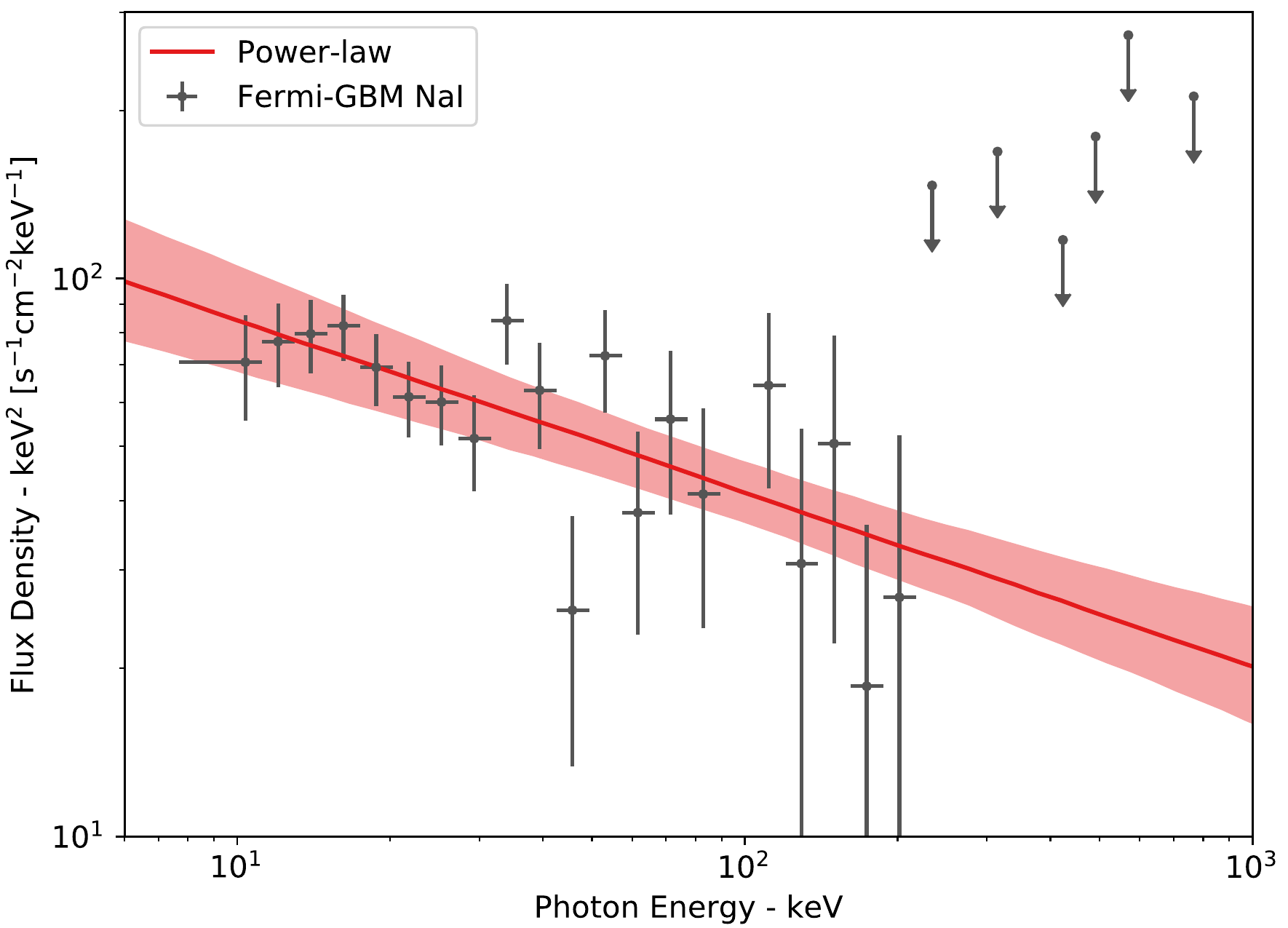}
\includegraphics[width=0.48\hsize,clip]{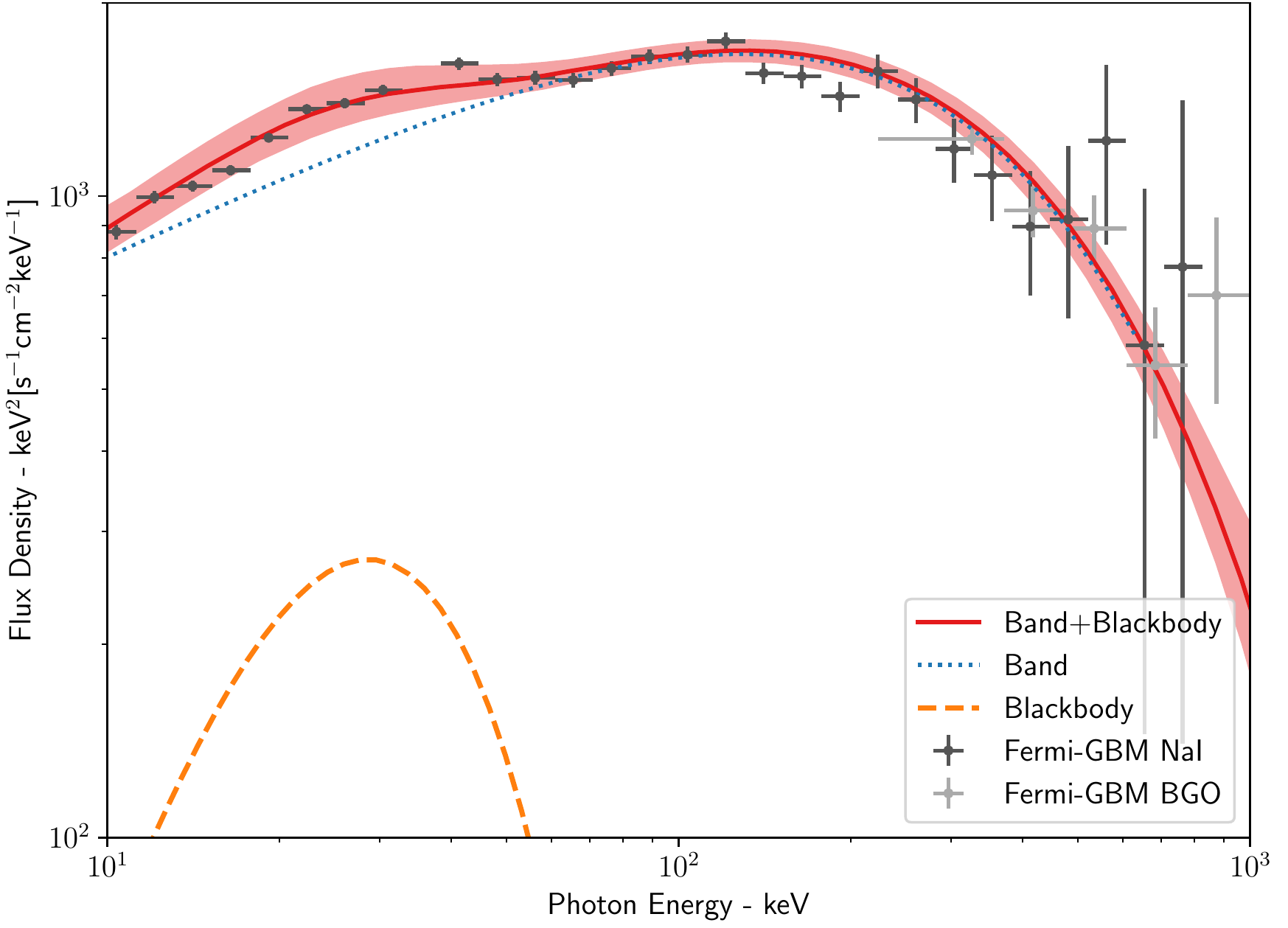}
\caption{\textbf{Left}: Spectrum of the precursor as observed by Fermi-GBM in the energy range of $8$--$900$~keV. The red line indicates the power-law fitting with power-law index $-2.31$, the red shadow is the $1$-sigma region. \textbf{Right}: Spectrum of the main prompt emission, from $8.72$~s to $12.30$~s. The dotted curve represents a band function with low-energy index $\alpha = -1.55$ and high-energy index $\beta = -3.48$; the peak energy is $E_p = 129$~keV. The additional blackbody component is represented by the orange dashed curve, with temperature $k T = 7.3$~keV. The composite fit with $1$-sigma confidence area is represented by the red line and red shadow.}
\label{fig:prompt_spectrum}
\end{figure*}

The first spike, the precursor, shows a power-law spectrum with a power-law index $-2.31\pm 0.08$ in its $2.75$~s duration, shown in Fig.~\ref{fig:prompt_spectrum} and in the appendix \ref{sec:model_comparison}. The averaged luminosity is $3.24^{+0.78}_{-0.55} \times 10^{49}$~erg~s$^{-1}$, and the integrated energy gives $7.98^{+1.92}_{-1.34}\times 10^{49}$~erg in the energy range from $1$~keV to $10$~MeV, the Friedmann-Lemaitre-Robertson-Walker metric with the cosmological parameters from Planck mission \citep{2018arXiv180706209P}\footnote{Hubble constant H0=($67.4\pm0.5$)~km/s/Mpc, matter density parameter $\Omega_M = 0.315\pm0.007$.} are applied on computing the cosmological distance throughout the whole paper.


\begin{table*}
\renewcommand\arraystretch{1.2}
\centering
\begin{tabular}{ccccc}
\hline\hline
\textbf{Time} & \textbf{Total Flux} & \textbf{Thermal Flux} & \textbf{Percentage} & \textbf{Temperature} \\
(s) & ($\text{erg}~ \text{s}^{-1} \text{cm}^{-2}$) & ($\text{erg}~ \text{s}^{-1} \text{cm}^{-2}$) &  & (keV) \\
\hline
8.72 - 10.80 & $5.6^{+1.1}_{-0.9} \times 10^{-6}$  & $4.1^{+3.2}_{-1.9} \times 10^{-7}$   & $7.3^{+5.8}_{-3.7} \%$      & $7.9^{+0.7}_{-0.7}$ \\
10.80 - 12.30 & $2.0^{+0.1}_{-0.1} \times 10^{-5}$     & $7.1^{+6.0}_{-3.3} \times 10^{-7}$      & $3.6^{+3.3}_{-1.6} \%$       & $5.6^{+0.5}_{-0.5}$ \\
\hline\hline
\end{tabular}
\caption{Parameters of the blackbody evolution in two time intervals. Parameters include the total flux, the thermal flux, the percentage of thermal flux and the temperature. One example of data fitting by Monte-Carlo iteration is shown in the appendix \ref{sec:data_fitting}. The time bin of $12.30~\text{s} - 22.54~\text{s}$ does not show convincing thermal component from the model comparison, still we report the fitting value, as a reference, from the cutoff power-law plus black body model, that the temperature is found as $2.1^{+0.5}_{-0.9}$ keV, the thermal flux is  $1.2^{+9.0}_{-1.1} \times 10^{-8} \text{erg}~ \text{s}^{-1} \text{cm}^{-2}$, and the total flux is $3.1^{+0.22}_{-0.2} \times 10^{-6} \text{erg}~ \text{s}^{-1} \text{cm}^{-2}$. }
\label{tab:thermalFlux}
\end{table*}

The second spike rises $10.29$~s after the starting time of the first spike ($8.72$~s since the trigger time), lasts $13.82$~s and emits $2.73^{+0.11}_{-0.10}\times 10^{51}$~erg in the $1$~keV--$10$~MeV energy band, i.e.~$84$ times more energetic than the first spike. The best fit of the spectrum is a Band function or a cutoff power-law, with an additional blackbody; see table \ref{tab:promptModel} in the appendix \ref{sec:model_comparison} for the model comparison of the time resolved analysis and figure~\ref{fig:prompt_spectrum} for the spectrum. We notice that the thermal component confidently exists in the second spike when the emission is luminous while, at times later than $12.30$~s, the confidence of the thermal component drops and a single cutoff power-law is enough to fit the spectrum. There could be many reasons for the missing thermal component at later times; for instance, the thermal component becomes less prominent and is covered by the non-thermal emission, or the thermal temperature cools to a value outside of the satellite energy band, or the thermal emission really disappears. In the present case the thermal blackbody component of temperature $\sim 7$~keV contributes $\sim 5\%$ to the total energy. 

From the evolution of the thermal spectrum and the parameters presented in table \ref{tab:thermalFlux}, it is possible to determine the velocity and the radius of the system in a model-independent way. Following \citet{2018ApJ...852...53R}, we obtain that the radius in each of the two time intervals is $1.4^{+0.6}_{-0.4}\times 10^{10}$~cm and $4.3^{+0.9}_{-0.6} \times 10^{10}$~cm respectively, and the expanding velocity is $0.53^{+0.18}_{-0.15}~c$. 

\subsection{Supernova}
\label{subsec:supernova}

\begin{figure}
\centering
\includegraphics[width=1.0\hsize,clip]{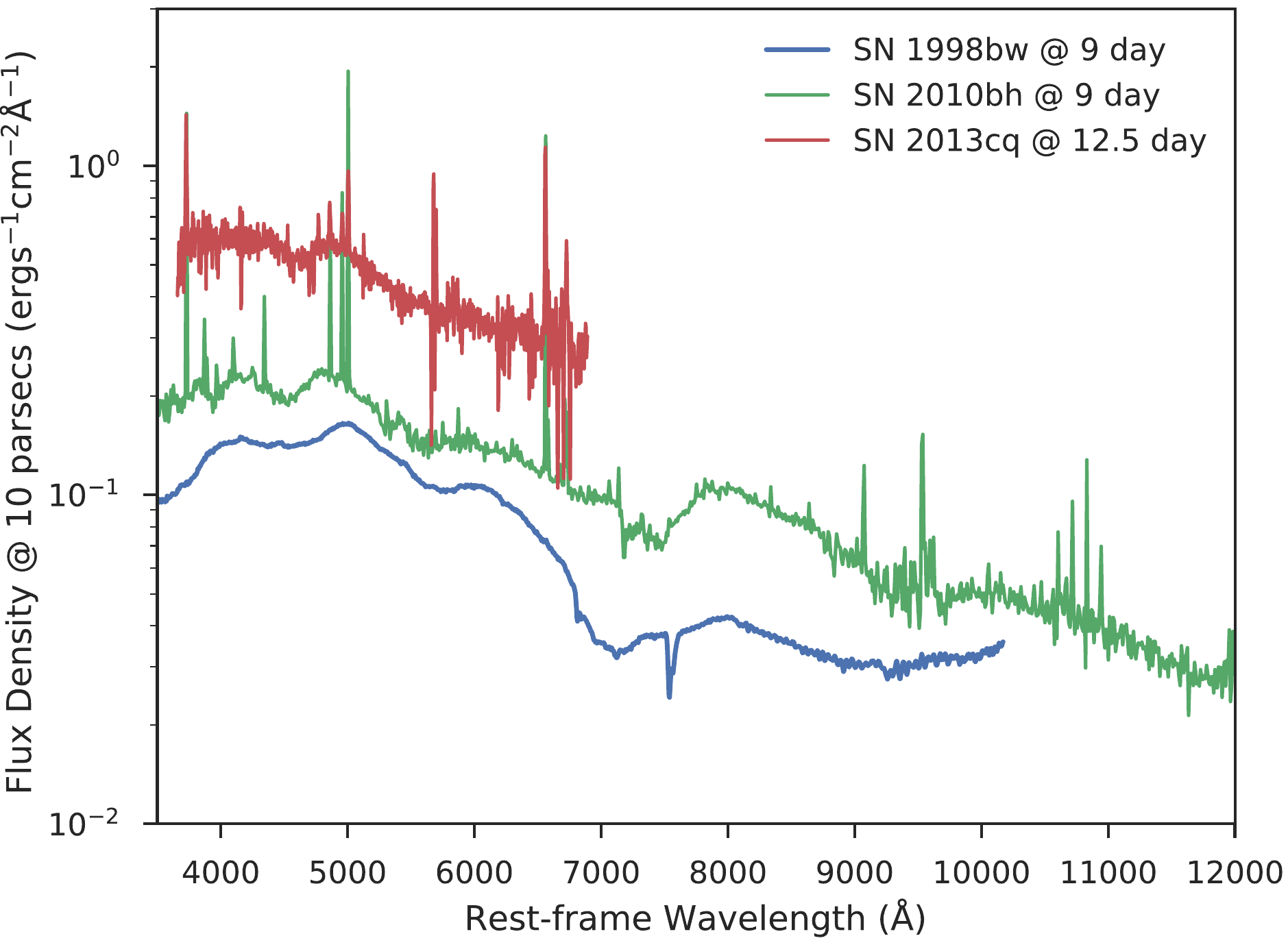}
\caption{Spectra comparison of three SNe: 1998bw, 2010bh, 2013cq, flux density is normalised at 10 parsec, data are retrieved from the Wiserep website (\url{https://wiserep.weizmann.ac.il}).}
\label{fig:sn2018fip}
\end{figure}

The optical signal of SN 2018fip associated with GRB 180728A was confirmed by the observations of the VLT telescope \citep{GCN23142,GCN23181}. The SN 2018fip is identified as a Type Ic SN, its spectrum at $\sim 8$ days after the peak of the optical light curve matches with the Type Ic SN 2002ap \citep{2002ApJ...572L..61M}, reported in \citet{2018TNSCR1249....1S}. In \citet{GCN23142}, there is the comparison of SN 2018fip with SN 1998bw and SN 2010bh, and in \citet{2013ApJ...776...98X}, there is the comparison of the SN associated with GRB 130427A, SN 2013cq, with SN 1998bw and SN 2010bh, associated with GRB 980425 \citep{1999A&AS..138..465G} and GRB 100316D \citep{2003ApJ...599L..95M}, respectively. We show in figure~\ref{fig:sn2018fip} the spectral comparison of SN 1998bw, SN 2010bh, and SN 2013cq.  We may conclude that the SNe are similar, regardless of the differences, e.g. in energetics ($\sim 10^{54}$~erg versus $\sim 10^{51}$~erg), of their associated GRBs (BdHN I versus BdHN II).



\section{Physical Interpretation} 
\label{sec:picture_of_physics}

All the observations in section~\ref{sec:picture_and_data} can be well interpreted within the picture of a binary system initially composing a massive CO$_{\rm core}$ and a NS.

\subsection{Prompt emission from a binary accretion system}

At a given time, the CO$_{\rm core}$ collapses forming a $\nu$NS at its center and producing a SN explosion. A strong shockwave is generated and emerges from the SN ejecta. A typical SN shockwave carries $\sim 10^{51}$~erg of kinetic energy \citep{1996snih.book.....A}, which is partially converted into electromagnetic emission by sweeping the circumburst medium (CBM) with an efficiency of $\sim 10\%$ \citep[see e.g.][]{2012SSRv..173..309B}. Therefore, the energy of $\sim 10^{50}$~erg is consistent with the total energy in the first spike. The electrons from the CBM are accelerated by the shockwave via the Fermi mechanism and emit synchrotron emission which explains the non-thermal emission with a power-law index $-2.31$ in the first spike.

The second spike with thermal component is a result of the SN ejecta accreting onto the companion NS. The distance of the binary separation can be estimated by the delay time between the two spikes, $\sim 10$~s. Since the outer shell of the SN ejecta moves at velocity $\sim 0.1~c$ \citep{2017AdAst2017E...5C}, we estimate a binary separation $\approx 3\times 10^{10}$~cm. Following \citet{2016ApJ...833..107B}, the total mass accreted by the companion NS gives $\sim 10^{-2}~M_{\odot}$, which produces an emission of total energy $\sim 10^{51}$~erg, considering the accretion efficiency as $\sim 10\%$ \citep{1992apa..book.....F}. The majority of the mass is accreted in $\sim 10$~s, with an accretion rate $\sim 10^{-3}~M_{\odot}$~s$^{-1}$, therefore, a spike with luminosity $\sim 10^{50}$~erg~s$^{-1}$ and duration $\sim 10$~s is produced, this estimation fits the second spike that observed well.

The time-resolved analysis of the blackbody components in the second spike indicate a mildly relativistic expanding source emitting thermal radiation. This emission is explained by the adiabatic expanding thermal outflow from the accretion region \citep{2006ApJ...646L.131F,2009ApJ...699..409F}. The Rayleigh-Taylor convective instability acts during the initial accretion phase driving material away from the NS with a final velocity of the order of the speed of light. This material expands and cools, by assuming the spherically symmetric expansion, to a temperature \citep{1996ApJ...460..801F,2016ApJ...833..107B} 
\begin{equation}
    T = 6.84\, \left(\frac{S}{2.85}\right)^{-1}\,\left( \frac{r}{10^{10}\, {\rm cm}} \right)^{-1}\,{\rm keV},
\end{equation}
where $S$ is the the entropy
\begin{multline}
    	S \approx 2.85 \left( \frac{M_{\rm NS}}{1.4\,M_\odot} \right)^{7/8}\left( \frac{\dot{M}_{\rm B}}{{10^{-3} M_\odot\,{\rm s}^{-1}}} \right)^{-1/4} \\ \times \left( \frac{r}{10^{10}\, {\rm cm}} \right)^{-3/8},
\end{multline}
in units of $k_B$ per nucleon. The system parameters in the above equations have been normalized to self-consistent values that fit the observational data, namely, the thermal emitter has a temperature $\sim 6$~keV, radius $\sim 10^{10}$~cm and expands with velocity $\sim 0.5~c$. 

To have more details of the time-resolved evolution: for the two time bins in table \ref{tab:thermalFlux}, an expanding speed of $0.53$c gives the radius $1.4 \times 10^{10}$ cm and $4.3 \times 10^{10}$ cm respectively, as we fitted from the data. The luminosities are $2.11 \times 10^{50}$ erg/s and $7.56 \times 10^{50}$ erg/s respectively. If assuming the accretion efficiency is $10\%$, from the luminosity we obtain the accretion rate as $1.18 \times 10^{-3}$ ~$M_{\odot}$~s$^{-1}$ and $4.32 \times 10^{-3}$ ~$M_{\odot}$~s$^{-1}$. By applying the above two equations, the theoretical temperature is obtained to be $5.83 \pm 1.25$ keV and $3.93 \pm 0.39$ keV, the the thermal flux are $1.22 \pm 0.97 \times 10^{-7}$ $\text{erg}~ \text{s}^{-1} \text{cm}^{-2}$ and $2.37 \pm 0.85 \times 10^-7$ $\text{erg}~ \text{s}^{-1} \text{cm}^{-2}$ respectively. If we assume the accretion efficiency is $7\%$, following the same procedure, the theoretical temperature shall be $8.32 \pm 1.78$ keV and $5.61\pm 0.56$ keV, the thermal flux shall be $5.78 \pm 3.89 \times 10^{-7}$ $\text{erg}~ \text{s}^{-1} \text{cm}^{-2}$ and $7.17 \pm 2.51 \times 10^{-7}$ $\text{erg}~ \text{s}^{-1} \text{cm}^{-2}$ respectively. The observed value in table \ref{tab:thermalFlux} are more consistent with the accretion efficiency of $7\%$.

The loss of rotational energy of the $\nu$NS, born after the SN explosion, powers the afterglow. This will be discussed in the next session.

\subsection{Afterglow from the newly born pulsar}
\label{sec:afterglow_from_pulsar}

\begin{figure*}
\centering
\includegraphics[width=0.8\hsize,clip]{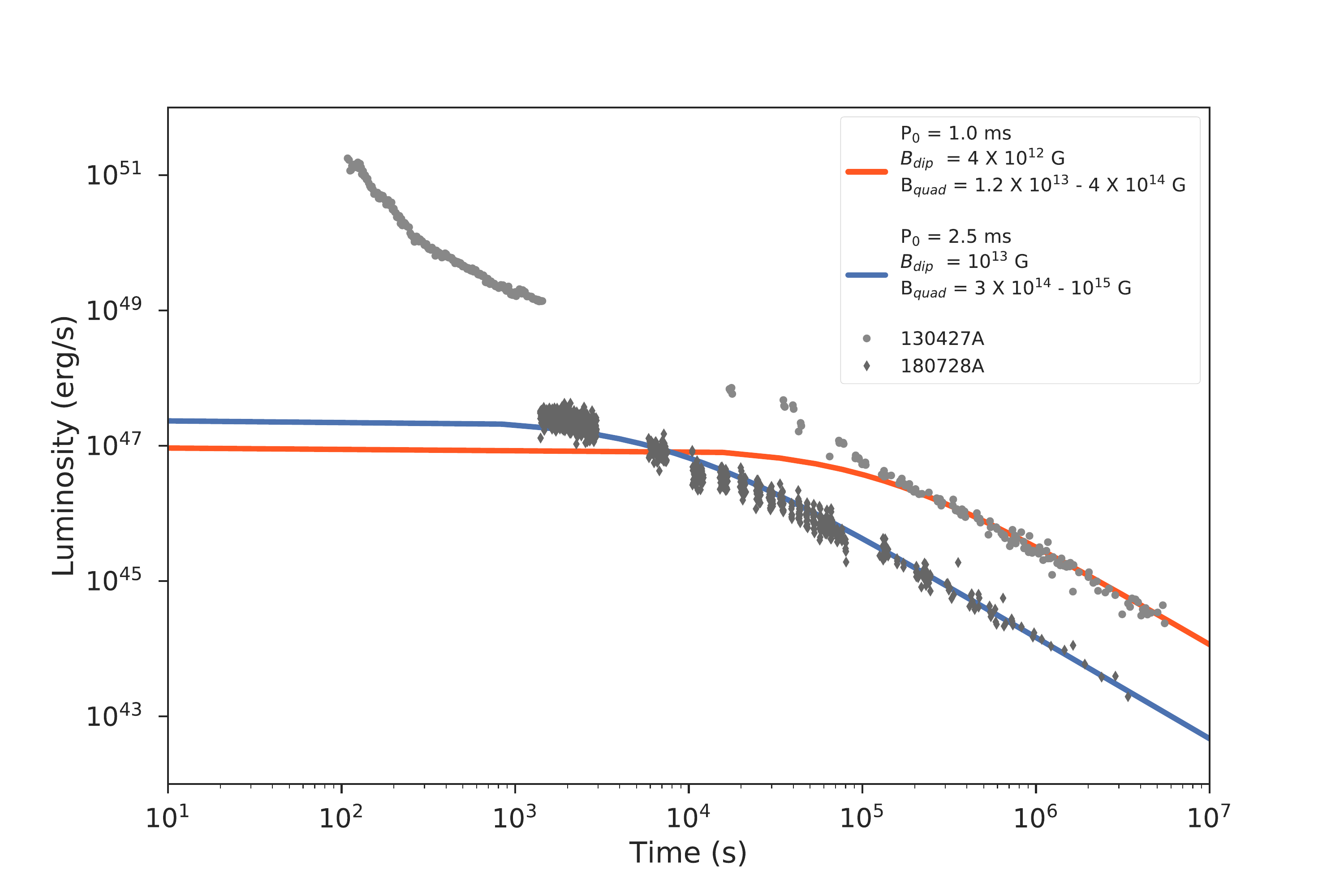}
\caption{Afterglow powered by the $\nu$NS pulsar: the grey and dark points correspond to the bolometric afterglow light curves of GRB 1340427A and 180728A, respectively.  The red and blue lines are the fitting of the energy injection from the rotational energy of the pulsar. The fitted parameters are shown in the legend, the quadruple field are given in a range, its upper value is $3.16$ times the lower value, this is due to the oscillation angle $\chi_2$, which is a free parameter. }
\label{fig:B4E13_2_5ms_luminosityb}
\end{figure*}

We have applied the synchrotron model of mildly-relativistic outflow powered by the rotational energy of the $\nu$NS to GRB 130427A \citep{2018ApJ...869..101R}. From it we have inferred a $1$~ms $\nu$NS pulsar emitting dipole and quadrupole radiation. Here we summarise this procedure and apply it to GRB 180827A. 

The late X-ray afterglow of GRB 180728A also shows a power-law decay of index $\sim -1.3$ which, as we show below, if powered by the pulsar implies the presence of a quadrupole magnetic field in addition to the traditional dipole one. The ``magnetar'' scenario with only a strong dipole field ($B_{dip} > 10^{14}$~G) is not capable to fit the late time afterglow \citep{1998A&A...333L..87D,2001ApJ...552L..35Z,2011MNRAS.413.2031M,2018ApJS..236...26L}. The dipole and quadrupole magnetic fields are adopted from \citet{2015MNRAS.450..714P}, where the magnetic field is cast into an expansion of vector spherical harmonics, each harmonic mode is defined by a set of the multipole order number $l$ and the azimuthal mode number $m$. The luminosity from a pure dipole ($l=1$) is
\begin{equation}
	L_{dip} = \frac{2}{3 c^3} \Omega^4 B_{dip}^2 R_{\rm NS}^6 \sin^2\chi_1,
\end{equation}
and a pure quadrupole ($l=2$) is
\begin{multline}
	L_{quad} = \frac{32}{135 c^5} \Omega^6 B_{quad}^2 R_{\rm NS}^8 \\ \times \sin^2\chi_1(\cos^2\chi_2+10\sin^2\chi_2),
\end{multline}
where $\chi_1$ and $\chi_2$ are the inclination angles of the magnetic moment, the different modes are easily separated by taking $\chi_1$ = 0 and any value of $\chi_2$ for $m = 0$, ($\chi_1$, $\chi_2$) = (90, 0) degrees for $m = 1$ and ($\chi_1$, $\chi_2$) = (90, 90) degrees for $m = 2$. 

The observed luminosity is assumed to be equal to the spin-down luminosity as
\begin{multline}
	\frac{dE}{dt} = -I \Omega  \dot{\Omega } =  - (L_{dip} + L_{quad}) \nonumber \\
    = - \frac{2}{3 c^3} \Omega^4 B_{dip}^2 R_{\rm NS}^6 \sin^2\chi_1  \left(1+\eta^2 \frac{16}{45} \frac{R_{\rm NS}^2 \Omega^2}{c^2}\right),
\end{multline}
and
\begin{equation}
    \eta^2 = (\cos^2\chi_2+10\sin^2\chi_2) \frac{B_{quad}^2}{B_{dip}^2}.
\label{eq:eta}
\end{equation}
where $I$ is the moment of inertia. The parameter $\eta$ relates to the ratio of quadrupole and dipole strength, $\eta = B_{quad}/B_{dip}$ for the $m=1$ mode, and $\eta = 3.16 \times B_{quad}/B_{dip}$ for the $m=2$ mode.

The bolometric luminosity is obtained by integrating the entire spectrum generated by the synchrotron model that fits the soft X-ray ($0.3$--$10$~keV) and the optical \citep[see][and figure 4 for example in]{2018ApJ...869..101R}. Approximately the bolometric luminosity has a factor of $\sim 5$ times more luminous than the soft X-ray emission. In figure \ref{fig:B4E13_2_5ms_luminosityb}, we show the bolometric luminosity light curve, the shape of the light curve is taken from the soft X-ray data since it offers the most complete time coverage. 

We assume that the bolometric luminosity required from the synchrotron model is equal to the energy loss of the pulsar. The numerical fitting result shows that the BdHN II of GRB 180728A forms a pulsar with initial spin $P_0 = 2.5$~ms, which is slower than the pulsar of $P_0 = 1$~ms pulsar from the BdHN I of GRB 130427A. Both sources have similar dipole magnetic field $10^{12}$--$10^{13}$~G and a quadrupole component $\sim 30$--$100$ stronger ($\eta = 100$) than the dipole one. The strong quadrupole field dominates the emission in the early years while the dipole radiation starts to be prominent later when the spin decays. This is because the quadrupole emission is more sensitive to the spin period, as $\propto \Omega^6$, while the dipole is $\propto \Omega^4$. Therefore, the $\nu$NS shows up a dipole behaviour when observed today, since the quadrupole dominates a very small fraction ($\lesssim 10^{-5}$) of the pulsar lifetime.




\section{A Consistent Picture and Visualisation} 
\label{sec:neutron_star_charactristic}

\begin{figure*}
    \centering
    \includegraphics[width=0.475\hsize,clip]{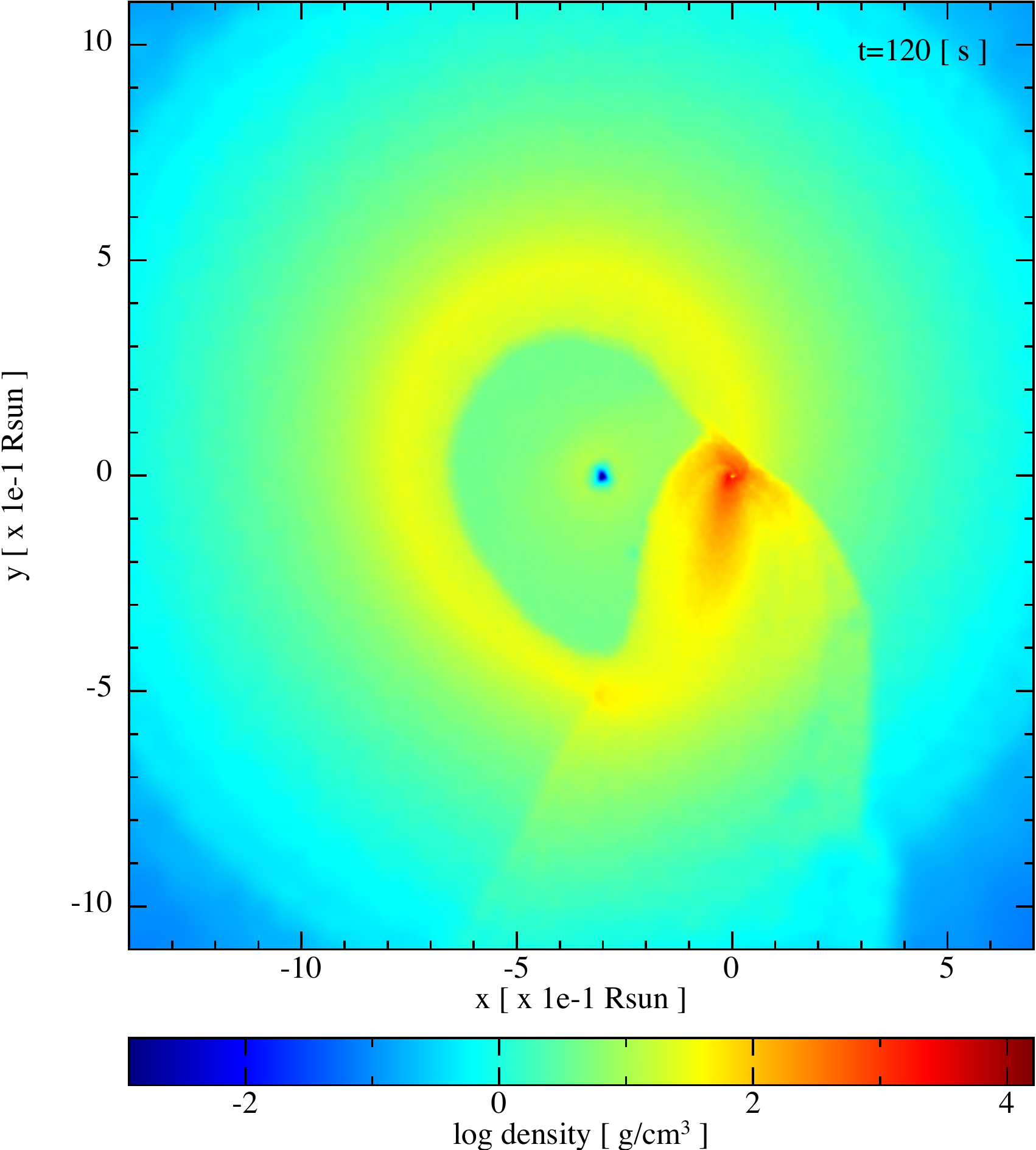}\includegraphics[width=0.49\hsize,clip]{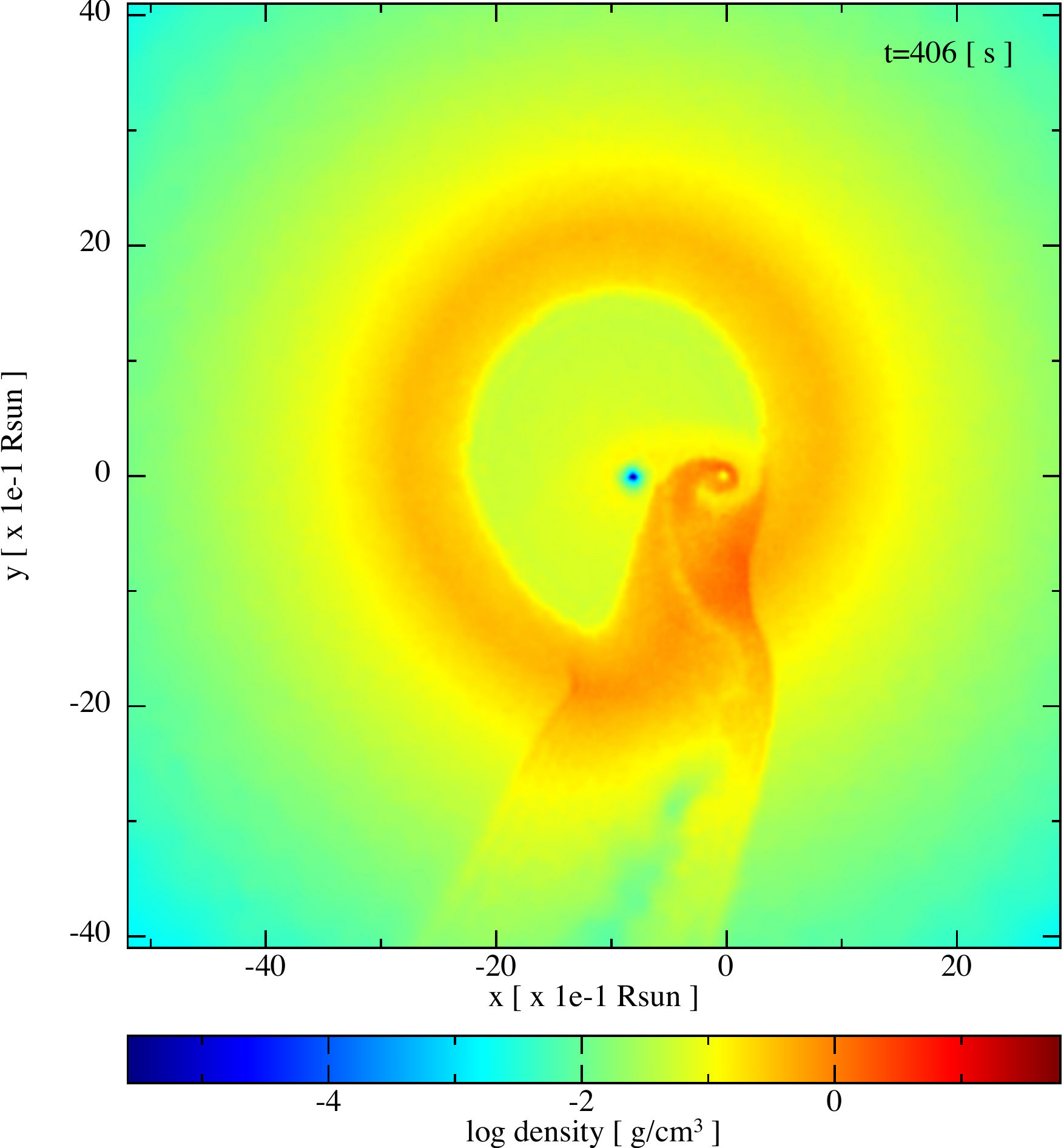}
    \caption{Two selected SPH simulations from \citet{2018arXiv180304356B} of the exploding CO$_{\rm core}$ as SN in presence of a companion NS: Model `25m1p08e' with $P_{\rm orb}=4.8$~min (left panel) and Model `25m3p1e' with $P_{\rm orb}=11.8$~min (right panel). The CO$_{\rm core}$ is taken from the $25~M_\odot$ ZAMS progenitor, so it has a mass $M_{\rm CO}=6.85~M_\odot$. The mass of the NS companion is $M_{\rm NS}=2~M_\odot$. The plots show the density profile on the equatorial orbital plane; the coordinate system has been rotated and translated in such a way that the NS companion is at the origin and the $\nu$NS is along the -x axis. The system in the left panel leads to a BdHN I and the snapshot is at the time of the gravitational collapse of the NS companion to a BH, $t=120$~s from the SN shock breakout ($t=0$ of our simulation). The system forms a new binary system composed by the $\nu$NS (at the center of the deep-blue region) and the BH formed by the collapsing NS companion (at the center of the red vortices). The system in the right panel leads to an BdHN II since the NS in this case does not reach the critical mass. This snapshot corresponds to $t=406$~s post SN shock breakout. In this simulation, the new system composed by the $\nu$NS and the NS companion becomes unbound after the explosion.}
    \label{fig:SPHsimulation}
\end{figure*}

In the previous sections, we have inferred the binary separation from the prompt emission, and the spin of the $\nu$NS from the afterglow data. In the following, we confirm the consistency of these findings by numerical simulations of these systems, and compare the commonalities and diversities of GRB 130427A and GRB 180728A as examples of BdHN I and BdHN II systems in our model, respectively.

From an observational point of view, GRB 130427A and GRB 180728A are both long GRBs, but they are very different in the energetic: GRB 130427A is one of the most energetic GRBs with isotropic energy more than $10^{54}$~erg, while GRB 180728A is in the order of $10^{51}$~erg, a thousand of times difference. GRB 130427A has observed the most significant ultra-high energy photons ($100$~MeV-- $100$~GeV, hereafter we call GeV photons), it has the longest duration ($>1000$~s) of GeV emission, and it has the highest energy of a photon ever observed from a GRB. In constrast, GRB 180728A has no GeV emission detected. As for the afterglow, the X-ray afterglow of GRB 130427A is more luminous than GRB 180728A, but they both share a power-law decaying index $\sim -1.3$ after $10^4$~s. After more than $10$~days, in both GRB sites emerges the coincident optical signal of a type Ic SN, and the SNe spectra are almost identical as shown in section~\ref{subsec:supernova}. 

BdHN I and II have the same kind of binary progenitor, a binary composed of a CO$_{\rm core}$ and a companion NS, but the binary separation/period is different, being larger/longer for BdHN II. 

The angular momentum conservation during the gravitational collapse of the pre-SN core that forms the $\nu$NS, i.e.~$J_{\rm CO} = J_{\nu \rm NS}$, implies that the latter should be fast rotating, i.e.: 
\begin{equation}
    \Omega_{\nu\rm NS} = \left(\frac{R_{\rm CO}}{R_{\nu\rm NS}}\right)^2\Omega_{\rm Fe}=\left(\frac{R_{\rm CO}}{R_{\nu\rm NS}}\right)^2\Omega_{\rm orb},
\end{equation}
where $\Omega_{\rm orb} = 2\pi/P_{\rm orb} = \sqrt{G M_{\rm tot}/a_{\rm orb}^3}$ from the Kepler law, being $M_{\rm tot}=M_{\rm CO}+M_{\rm NS}$ the total mass of the binary before the SN explosion, and $M_{\rm CO} = M_{\nu\rm NS}+M_{\rm ej}$. We have assumed that the mass of the $\nu$NS is set by the mass of the iron core of the pre-SN CO$_{\rm core}$ and that it has a rotation period equal to the orbital period owing to tidal synchronization.

From the above we can see that $\nu$NS rotation period, $P_{\nu\rm NS}$ has a linear dependence with the orbital period, $P_{\rm orb}$. Therefore, the solution we have obtained for the rotation period of the $\nu$NS born in GRB 130427A ($P_{\nu\rm NS} \approx 1$~ms) and in the GRB 180728A ($P_{\nu\rm NS} \approx 2.5$~ms), see Fig.~\ref{fig:B4E13_2_5ms_luminosityb}, implies that the orbital period of the BdHN I would be a factor $\approx 2.5$ shorter than the one of the BdHN II. Based on this information, we seek for two systems in our simulations presented in \citet{2018arXiv180304356B} with the following properties: the same (or nearly) SN explosion energy, same pre-SN CO$_{\rm core}$ and initial NS companion mass, but different orbital periods, i.e. $P_{\rm II}/P_{\rm I}\approx 2.5$. The more compact binary leads to the BdHN I and the less compact one to the BdHN II and, by angular momentum conservation, they lead to the abovementioned $\nu$NSs.

We examine the results of the simulations for the pre-SN core of a $25~M_\odot$ zero-age main-sequence (ZAMS) progenitor and the initial mass of the NS companion $M_{\rm NS}=2~M_\odot$. A close look at Tables 2 and 7 in \citet{2018arXiv180304356B} show that, indeed, Model `25m1p08e' with $P_{\rm orb} = 4.81$~min ($a_{\rm orb}\approx 1.35\times 10^{10}$~cm) and Model `25m3p1e' with $P_{\rm orb} = 11.8$~min ($a_{\rm orb}\approx 2.61\times 10^{10}$~cm) give a consistent solution. In the Model `25m1p08e' the NS companion reaches the critical mass (secular axisymmetric instability) and collapses to a BH; this model produces a BdHN I. In the Model `25m3p1e' the NS companion does not reach the critical mass; this system produces an BdHN II. The system leading to the BdHN I remains bound after the explosion while, the one leading to the BdHN II, is disrupted. Concerning the $\nu$NS rotation period, adopting $R_{\rm CO}\sim 2.141\times 10^8$~cm \citep[see Table~1 in][]{2018arXiv180304356B}, $P_{\rm orb}\sim 4.81$~min and $P_{\rm orb}\sim 11.8$~min leads to $P_{\nu\rm NS} \sim 1$~ms and $2.45$~ms, respectively. We show in Fig.~\ref{fig:SPHsimulation} snapshots of the two simulations.
%

\section{Conclusion}
\label{sec:conclusion}

The classification of GRBs in nine different subclasses allows us to identify the origin of a new GRB with known redshift from the observation of its evolution in the first hundred seconds. Then, we are able to predict the presence of an associated SN in the BdHN and its occurring time. We reviewed our previous successful prediction of a BdHN I in our model: GRB 130427A/SN 2013cq, and in this article, we presented our recent successful prediction of a BdHN II in our model: GRB 180728A/SN 2018fip.

The detailed observational data of GRB 180728A, for the first time, allowed us to follow the evolution of a BdHN II. The collapse of CO$_{\rm core}$ leads to a SN. We determine that the corresponding shockwave with energy $\sim 10^{51}$~erg emerges and produces the first $2$~s spike in the prompt emission. The SN ejecta expands and reaches at $\sim 3\times 10^{10}$~cm away from the NS companion. The accretion process starts with a rate $\sim 10^{-3}~M_{\odot}$~s$^{-1}$, the second powerful spike lasting $10$~s with luminosity $\sim 10^{50}$~erg~s$^{-1}$, and a thermal component at temperature $\sim 7$~keV.


A $\nu$NS is formed from the SN. The role of $\nu$NS powering the afterglow has been evidenced in our study of GRB 130427A \citep{2018ApJ...869..101R}. This article emphasises its application on GRB 180728A. The $\nu$NS pulsar loses its rotational energy by dipole and quadrupole emission. In order to fit the observed afterglow data using a synchrotron model \citep{2018ApJ...869..101R}, we require an initial $1$~ms spin pulsar for GRB 130427A, and a slower spin of $2.5$~ms for GRB 180728A. For close binary systems, the binary components are synchronised with the orbital period, from which we are able to obtain the orbital separation by inferring the CO$_{\rm core}$ period from the $\nu$NS one via angular momentum conservation. This second independent method leads to a value of the binary separation in remarkable agreement with the one inferred from the prompt emission, which shows the self-consistency of this picture. The SNe spectra observed in BdNH I and in BdHN II are similar, although the associated two GRBs markedly differ in energy. The SN acts as a catalyst; it triggers the GRB process. After losing a part of the ejecta mass by hypercritical accretion, the remaining SN ejecta are heated by the GRB emission, but the nuclear composition, which relates to the observed optical emission owing to the nuclear decay of nickel and cobalt \citep{1996snih.book.....A} is not influenced by such a GRB-SN interaction. 

Besides providing the theoretical support of the BdHN I and II realisation, we have presented 3D SPH simulations that help in visualising the systems (see figure~\ref{fig:SPHsimulation}).

In short, we made a successful prediction of SN 2018fip associated with GRB 180728A based on our GRB classification that GRB 180728A belongs to BdHN II. The observations of the prompt emission and the afterglow portray, for the first time, a complete transitional stage of two binary stars. We emphasise the $\nu$NS from supernova playing a dominant role in the later afterglow, the comparison to GRB 130427A, a typical BdHN I in our model, is demonstrated and visualised. 

\acknowledgments

The confirmation of the SN appearance, as well as the majority of this work were performed during the R.R. and Y.W.'s visit to the \textit{Yau Mathematical Sciences Center} in Tsinghua University, Beijing. We greatly appreciate the kind hospitality of and the helpful discussion with Prof. Shing-Tung Yau. We also acknowledge Dr. Luca Izzo for discussions on the SNe treated in this work. We thank to the referee for the constructive comments that helped clarify many concepts and strengthen the time-resolved analysis.

\bibliographystyle{aasjournal}
\bibliography{180728A}

\appendix

\section{GCNs}
\label{sec:gcns}

\begin{displayquote}
\textbf{GCN 14526} - 
\textit{\textbf{GRB 130427A: Prediction of supernova appearance}}\\
\textit{The late x ray observations of GRB 130427A by Swift-XRT clearly evidence a pattern typical of a family of GRBs associated to supernova (SN) following the Induce Gravitational Collapse (IGC) paradigm \citep{2012ApJ...758L...7R,2013A&A...552L...5P}. We assume that the luminosity of the possible SN associated to GRB 130427A would be the one of 1998bw, as found in the IGC sample described in \citep{2013A&A...552L...5P}. Assuming the intergalactic absorption in the I-band (which corresponds to the R-band rest-frame) and the intrinsic one, assuming a Milky Way type for the host galaxy, we obtain a magnitude expected for the peak of the SN of I = 22 - 23 occurring 13-15 days after the GRB trigger, namely between the 10th and the 12th of May 2013. 
Further optical and radio observations are encouraged.}
\end{displayquote}

\begin{displayquote}
\textit{\textbf{GCN 23066 -  GRB 180728A: A long GRB of the X-ray flash (XRF) subclass, expecting supernova appearance}}

\textit{GRB 180728A has $T_{90}=6.4$~s \citep{GCN23055}, peak energy 142 (-15,+20)~keV, and isotropic energy $E_{\rm iso} = (2.33 \pm 0.10) \times 10^{51}$~erg \citep{GCN23061}. It presents the typical characteristic of a subclass of long GRBs called X-ray flashes\footnote{The previous name of BdHN I} \citep[XRFs, see][]{2016ApJ...832..136R}, originating from a tight binary of a CO$_{\rm core}$ undergoing a supernova explosion in presence of a companion neutron star (NS) that hypercritically accretes part of the supernova matter. The outcome is a new binary composed by a more massive NS (MNS) and a newly born NS ($\nu$NS). Using the averaged observed value of the optical peak time of supernova \citep{2017AdAst2017E...5C}, and considering the redshift $z=0.117$ \citep{GCN23055}, a bright optical signal will peak at $14.7 \pm 2.9$ days after the trigger (12 August 2018, uncertainty from August 9th to August 15th) at the location of RA=253.56472 and DEC=-54.04451, with an uncertainty $0.43$~arcsec \citep{GCN23064}. The follow-up observations, especially the optical bands for the SN, as well as attention to binary NS pulsar behaviours in the X-ray afterglow emission, are recommended.}
\end{displayquote}

\begin{displayquote}
\textit{\textbf{GCN 23142 - GRB 180728A: discovery of the associated supernova}}

\textit{... Up to now, we have observed at three epochs, specifically at 6.27, 9.32 and 12.28 days after the GRB trigger. The optical counterpart is visible in all epochs using the X-shooter acquisition camera in the g, r and z filters. We report a rebrightening of 0.5 $\pm$ 0.1 mag in the r band between 6.27 and 12.28 days. This is consistent with what is observed in many other lo 170827w-redshift GRBs, which in those cases is indicative of an emerging type Ic SN ...}
\end{displayquote}

\section{Data Fitting}
\label{sec:data_fitting}

\begin{figure*}
    \centering
    \includegraphics[width=\hsize,clip]{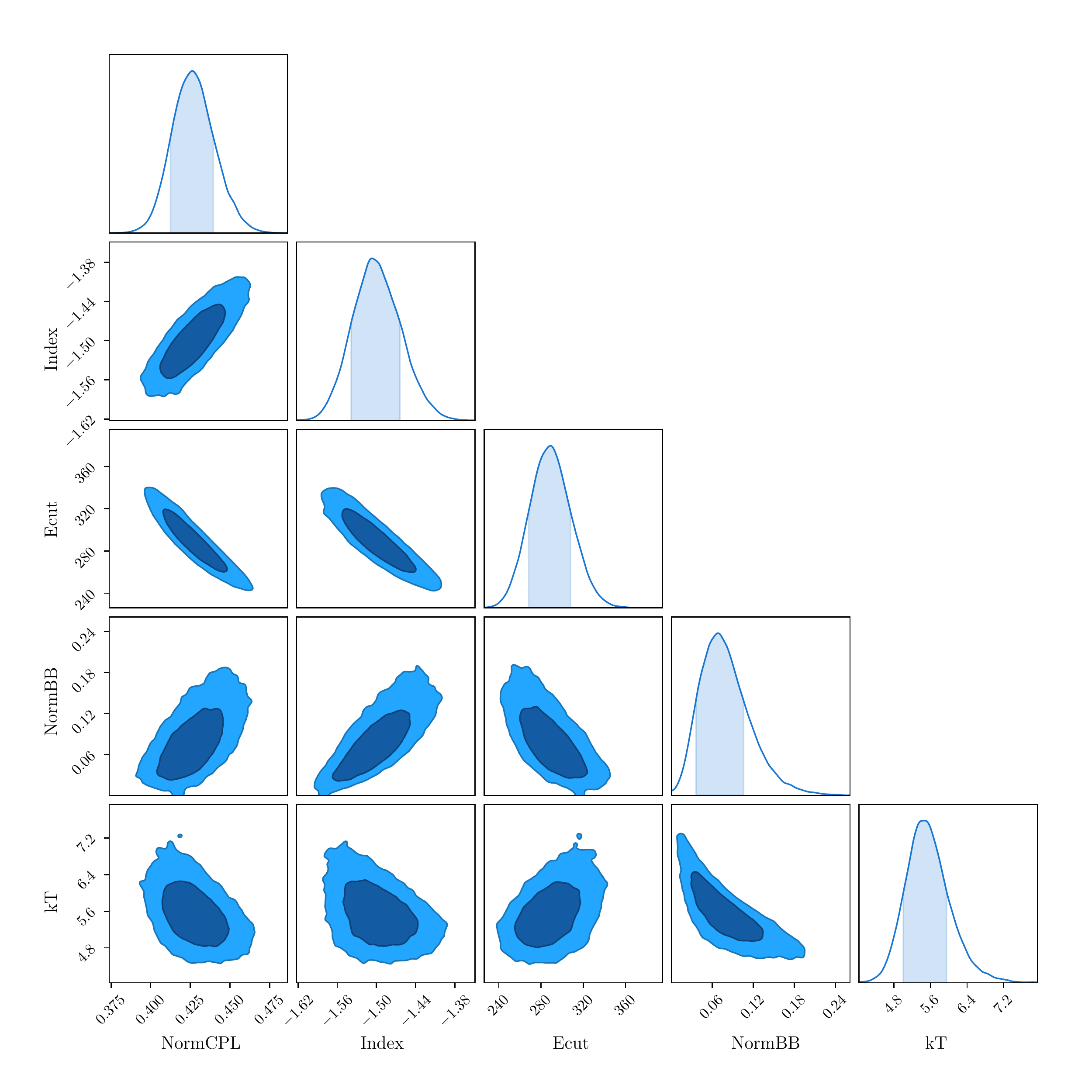}
    \caption{An example of fitting the Fermi-GBM data from $10.80$~s to $12.30$~s. We apply $20$ chains, each chain iterates $10^4$ times and burns the first $10^3$ times. The parameters are  normalisation (NormCPL), cut-off energy (ECut) and power-law index (Index) of the cut-off power-law model, as well as  normalisation (NormBB) and temperature ($k$T) of the blackbody model.}
    \label{fig:datafitting}
\end{figure*}

Data are fitted by applying the Monte Carlo Bayesian iterations using a Python package: The Multi-Mission Maximum Likelihood framework (3ML) \footnote{\url{https://github.com/giacomov/3ML}}. An example is shown in figure \ref{fig:datafitting}.

\section{Model Comparison}
\label{sec:model_comparison}

Spectra are fitted by Bayesian iterations. The AIC is preferred for comparing non-nested models, and BIC is preferred for nested models \citep{10.2307/2291091}. Log(likelihood) is adopted by the method of maximum likelihood ratio test which is treated as a reference of the model comparison \citep{10.2307/1912557}.  Parameters are shown in table \ref{tab:promptModel}.

\begin{table*}
\begin{center}
\begin{tabular}{lllccc}
\hline\hline
\textbf{Segment}    & \textbf{Time (s)} & \textbf{Model} & \textbf{Log(Likelihood)} & \textbf{AIC} & \textbf{BIC} \\
\hline
Spike 1             & -1.57 - 1.18      & {\ul PL}       & 430.03                   & 864.17       & 869.59       \\
Precursor           &                   & CPL            & 430.03                   & 866.27       & 874.35       \\
(NaI7)              &                   & Band           & 429.72                   & 867.80       & 878.49       \\
                    &                   & PL+BB          & 429.77                   & 867.90       & 878.59       \\
                    &                   & CPL+BB         & 429.77                   & 870.08       & 883.36       \\
                    &                   & Band+BB        & 429.61                   & 871.98       & 887.79       \\
\hline
Spike 2             & 8.72 - 10.80      & PL             & 947.20                   & 1898.46      & 1905.33      \\
(NaI7+BGO1)         &                   & CPL            & 838.91                   & 1685.93      & 1696.22      \\
                    &                   & Band           & 831.02                   & 1670.21      & 1683.90      \\
                    &                   & PL+BB          & 947.21                   & 1902.59      & 1916.27      \\
                    &                   & CPL+BB         & 827.67                   & 1665.60      & 1682.66      \\
                    &                   & {\ul Band+BB}  & 823.90                   & 1660.17      & 1680.59      \\ \cline{2-6}
                    & 10.80 - 12.30     & PL             & 1334.10                  & 2672.25      & 2679.13      \\
                    &                   & CPL            & 809.83                   & 1625.76      & 1636.05      \\
                    &                   & Band           & 821.25                   & 1650.68      & 1664.37      \\
                    &                   & PL+BB          & 1334.10                  & 2676.38      & 2690.06      \\
                    &                   & {\ul CPL+BB}   & 794.79                   & 1599.85      & 1616.91      \\ 
                    &                   & Band+BB        & 794.80                   & 1599.86      & 1616.92      \\ \cline{2-6}
                    & 12.30 - 22.54     & PL             & 1366.08                  & 2736.23      & 2742.79      \\
                    &                   & {\ul CPL}      & 1216.52                  & 2439.16      & 2448.97      \\
                    &                   & Band           & 1366.43                  & 2741.06      & 2754.09      \\
                    &                   & PL+BB          & 1366.08                  & 2740.37      & 2753.40      \\
                    &                   & CPL+BB         & 1215.52                  & 2443.35      & 2459.58      \\
                    &                   & Band+BB        & 1366.63                  & 2745.69      & 2765.11      \\
\hline\hline
\end{tabular}
\end{center}
\caption{Model comparison of time-resolved analysis of Fermi-GBM data. In the segment column, the name and the instruments are presented. The time column gives the time interval. In the model column, the model with an underline is the preferred one. We have used the following abbreviations: PL (Power-law), CPL (cutoff power-law) and BB (blackbody).  }
\label{tab:promptModel}
\end{table*}

\end{document}